\documentstyle[preprint,aps,epsf]{revtex}

\begin{document}

\draft

\preprint{HIP-1998-29/TH}

\title{Radiative symmetry breaking and the ${b \rightarrow s \gamma}$ decay in 
generalized GMSB models}

\author{H.~Hamidian$^{\rm a}$, K.~Huitu$^{\rm b}$,
K.~Puolam\"{a}ki$^{\rm b}$, and D.-X.~Zhang$^{\rm b}$}

\address{$^{\text{a}}$Department of Physics, Stockholm University, Box
6730, S-113~85 Stockholm, Sweden}

\address{$^{\text{b}}$Helsinki Institute of Physics, P.O.Box 9,
FIN-00014 University of Helsinki, Finland}


\maketitle

\begin{abstract}

We study a class of generalized models of gauge mediated supersymmetry
breaking (GMSB). We find the parameters and the full particle spectrum
of the minimal supersymmetric standard model (MSSM) for all GMSB
models with messenger multiplicities that satisfy the perturbativity
of the gauge couplings up to the GUT scale. We give a detailed
description of the algorithm that has been used to predict all the
unmeasured parameters of the MSSM by taking (one-loop) radiative
effects into account. We also calculate the branching ratio $BR(b
\rightarrow s \gamma )$ and find that it always turns out to be larger
than the standard model prediction. However, we find that the
branching ratio typically remains within the current experimental
bounds, except for some special cases with a light charged Higgs
boson, or a small supersymmetry breaking scale and a negative
$\mu$-parameter.

\end{abstract}


A very important question in all supersymmetric (SUSY) grand unified
theories beyond the standard model (SM) is to understand the mechanism
of SUSY breaking and its communication to the observable (low-energy)
sector. Although the answer to this question is still far from clear,
there are a number of criteria which must be met
regardless of any particular mechanism that breaks SUSY. For example,
the so-called soft SUSY breaking mass terms must be small enough to
ensure the stability of the Higgs mass against a large hierarchy of
scales, while they must be large enough to have evaded the
experimental searches so far.

SUSY breaking can be communicated to the visible sector either by
gravitational interactions, as in supergravity (SUGRA) inspired
models, or by SM gauge interactions, as in theories with
gauge-mediated SUSY breaking (GMSB), which were initially studied
nearly two decades ago \cite{gmsb}.  The former models naturally lead
to a universal soft-breaking sector near the Planck scale but are
incapable of sufficiently suppressing large flavor violations at
observable scales. Over the years many attempts have been made to
solve this notorious SUSY flavor problem by invoking the ``super-GIM''
mechanism by assuming mass-degeneracy among the squarks and sleptons
of a given flavor \cite{donoghue83}.  However, it is now understood
that a number of rather generic phenomena, including non-minimal
K\"{a}hler potentials, GUT effects, and superstring thresholds, can
break degeneracy in SUGRA models and, in fact, even if all these could
be somehow ruled out, scalar mass degeneracies will tend to be spoiled
since flavor physics occurs below the scale at which SUSY-breaking is
communicated to the SM. On the other hand, in GMSB theories gauge
interactions provide flavor-symmetric SUSY breaking terms and thus
naturally suppress the flavor-changing neutral currents associated
with (universal) soft squark and slepton masses. Furthermore, due to
the relatively low value of the SUSY-breaking scale in GMSB theories,
this universality remains nearly unbroken as one evolves the mass parameters
to the electroweak scale by using the renormalization group equations
(RGEs). Among other attractive features of GMSB theories are the small
number of free parameters, compared to the minimal supersymmetric
standard model (MSSM), and the possibility of providing a solution to
the SUSY CP problem. These features, together with the recent advances
in understanding nonperturbative effects in SUSY gauge theories
\cite{seiberg1} and the discovery of many new mechanisms for dynamical
SUSY breaking (DSB) \cite{skiba}, have led to a revival of interest in
GMSB models \cite{dineetal}.

In the minimal version of the GMSB model \cite{dineetal} the messenger
fields which communicate SUSY breaking to the visible sector belong to
the ${\bf 5}+\overline {\bf 5}$ or ${\bf 10}+\overline {\bf 10}$
representations of the $SU(5)$ gauge group, and the messenger Yukawa
couplings in any given $SU(5)$ representation are taken to be equal at
the unification scale $M_{\rm GUT}$. In order to communicate SUSY
breaking, in addition to the particles in the minimal supersymmetric
standard model (MSSM), the minimal GMSB theory contains at least one
singlet superfield $S$ which couples to vector-like messenger
superfields $V+\overline V$ through the superpotential interaction
\begin{eqnarray}
W_{\rm mess}=\lambda_V SV\overline V.
\label{eq1}
\end{eqnarray}
At a scale $\Lambda_{\text{M}}\sim 10^2-10^4$ TeV, SUSY is broken and 
both the lowest and
$F$-component, $F_S$, of the singlet superfield $S$ acquire vacuum
expectation values (VEVs) through their interactions with the hidden
sector.  The VEV $\langle S \rangle$ gives masses to the vector-like
supermultiplets $V+\overline V$, while $\langle F_S \rangle$ induces
mass splitting within the supermultiplets. Consequently, the gaugino
and sfermion masses are generated through their gauge couplings to the
messenger fields. The gauginos receive masses at one-loop,
$m_{\lambda} \sim (\alpha / 4\pi) \Lambda_{\text{SUSY}}$, where
$\Lambda_{\text{SUSY}}=\langle F_S \rangle/\langle S \rangle$, while
squarks and sleptons do so only at two-loop order, ${\tilde m}^2 \sim
(\alpha / 4\pi)^2 \Lambda_{\text{SUSY}}^2$. This implies that
$m_{\lambda} \sim {\tilde m}$, which is another attractive feature
of GMSB theories.

In the non-minimal generalizations of GMSB models \cite{martin} the
messenger fields do not necessarily form complete $SU(5)$ GUT
multiplets and one is naturally led to consider messenger fields which
belong to incomplete representations of the $SU(5)$ gauge group. This
can be seen by noting that the unification of the messenger Yukawa
couplings at the GUT scale---whose MSSM analogue is the so-called
$b-\tau$ unification \cite{chanowitz}---is not necessarily required
for gauge unification. For example, suppose that in addition to $S$
there exist singlet superfields, $S'$, whose VEVs (but not the VEVs of
their $F$-components) are just below the GUT scale, and which couple
only to some components of the $SU(5)$ multiplet. Then, within the
$SU(5)$ multiplet, these superfields acquire masses of order ${\cal
O}(M_{\rm GUT})$ and decouple from the low-energy spectrum. The other
components, which get their masses only through couplings with the
superfield $S$, obtain masses of order $\lambda \langle S \rangle \sim
\Lambda_{\text{M}}$. Since $\sqrt{\langle F_S \rangle}$ is much
smaller than the masses of the heavy superfields, these (missing)
particles make negligible mass contributions and play a less important
role in determining the MSSM mass spectrum.

Although the phenomenological aspects of the minimal GMSB theories
have been studied extensively \cite{gmsbph,borzu,carone}, the 
generalized GMSB
models---which may be more relevant for the construction of
realistic SUSY GUTs---have been much less investigated. This is partly
due to the larger messenger field content in the generalized GMSB
theories which makes the study of these models technically more
demanding. In a previous work \cite{hamidian} we have studied the
phenomenological implications of generalized supersymmetric $SU(5)$
GUTs with GMSB by calculating the upper limits on $\tan \beta$ from
nucleon decay and have found that the predicted values of $\tan \beta$
are mostly inconsistent with the constraints from nucleon decay.  Our
results suggest that in order to construct phenomenologically viable
models, more complicated scenarios at the unification scale have to be
considered.

In this Letter, we do not wish to speculate further on GUT scale
physics in relation to GMSB theories, but instead take a bottom-up
approach. The outline of our investigation is as follows: by assuming
the most general sector of chiral messenger superfields, and
restricting the number of messenger fields by demanding that the gauge
couplings remain perturbative \cite{martin}, 32 distinct possibilities
for the messenger multiplicities are found.  We assume vanishing
bilinear and trilinear couplings in the soft scalar potential at the
messenger scale.  For each messenger sector there can be up to three
different MSSM particle spectra and we list the full particle spectrum
of all consistent models. In all the cases $\tan \beta$ is large and
the $\mu$-term can be either positive or negative, depending on the
details of the messenger field spectrum. Our results are based on the
calculation of the one-loop effective potential for a given messenger
sector and we use the SM renormalization group equations (RGEs) to
one-loop order to run the gauge and Yukawa couplings from the
electroweak scale, $m_{\rm Z}$, to the squark mass scale, $m_{\rm
SUSY}$. 

We then compute the branching ratios for the $b \rightarrow s \gamma$
decay and compare it to the SM prediction. This decay channel is very
sensitive to new physics in the large $\tan \beta$ regime. The
dominant non-SM contributions arise through constructive interference
with the charged Higgs-boson loops and the next most important contributions
are due to charginos. We find that in all the representative cases the
branching ratios for the $b \rightarrow s \gamma$ decay are larger
than the SM prediction.
For the minimal GMSB models our results qualitatively agree with
Borzumati's in \cite{borzu}.

Following Martin \cite{martin}, we specify the messenger
superfields in the generalized GMSB models as follows:
\begin{eqnarray}
n_Q:~~~~ && Q+{\overline Q}=({\bf 3},{\bf 2},\frac{1}{6}) +
\text{conj.}, \nonumber \\ n_U:~~~~ && U+{\overline U}=({\overline
{\bf 3}},{\bf 1},-\frac{2}{3}) + \text{conj.},\nonumber\\ n_D:~~~~ &&
D+{\overline D}=({\overline {\bf 3}},{\bf 1},\frac{1}{3}) +
\text{conj.},\nonumber\\ n_L:~~~~ && L+{\overline L}=({\bf 1},{\bf
2},-\frac{1}{2}) + \text{conj.},\nonumber\\ n_E:~~~~ && E+{\overline
E}=({\bf 1},{\bf 1},1) + \text{conj.} ,
\label{eq:mfields}
\end{eqnarray}
where the multiplicities of the messenger fields are denoted by
$(n_Q,n_U,n_D,n_L,n_E)$.

At this stage, several restrictions can be used to reduce the number
of possible models (see Ref.~\cite{martin} for further discussion). By
requiring that the gauge couplings remain perturbative, that the
messenger field masses taking part in SUSY breaking not greatly exceed
$10^4$ TeV, and that all the gauginos acquire a non-vanishing mass, one ends
up with 53 models obeying the following costraints on the messenger
multiplicities \cite{martin}:
\begin{eqnarray}
(n_Q,n_U,n_D,n_L,n_E) \leq & (1,0,2,1,2)\nonumber\\ \text{or}~ &
(1,1,1,1,1)\nonumber\\ \text{or}~ & (1,2,0,1,0)\nonumber\\ \text{or}~ &
(0,0,4,4,0).
\label{eq:martinlimits}
\end{eqnarray}
Some of these 53 models give equivalent MSSM physics, since the
sparticle masses are unaffected by changing $Q+\bar Q+E+\bar E
\leftrightarrow 3(D+\bar D+L+\bar L)$ (equivalent to ${\bf 10}+{\overline{\bf
10}} \leftrightarrow 3({\bf 5}+{\overline{\bf 5}})$) or $U+\bar U
\leftrightarrow D+\bar D+E+\bar E$. Taking these permutation symmetries into
account one is finally left with 32 different possibilities.

The vanishing trilinear and bilinear couplings in the soft
scalar potential at the messenger scale makes GMSB
models phenomenologically extremely predictive and fixes all the (as
yet unmeasured) parameters of the MSSM for a given messenger sector.
However, in order to implement this property one must compute the
(radiatively corrected) effective potential and search for the
minimum.  We shall now proceed with the study of these radiative
effects in the 32 models identified above and discuss their
implications.

Our goal is to determine all the MSSM parameters for any given messenger
sector. We specify the messenger sectors by their corresponding messenger 
multiplicities
in (\ref{eq:mfields}) and by the messenger scale $\Lambda_{\rm M}
\simeq \lambda \langle S \rangle$. We have for simplicity taken only
one effective messenger scale; all messengers have approximately the
same mass. Generalization to multiple messenger scales would be
straightforward.

For a given messenger sector, and at the messenger scale, we can calculate 
the bilinear term $B$ (which is assumed to vanish) as a function
of $\tan \beta=\langle H^0_u \rangle / \langle H^0_d \rangle$ and the
sign of the $\mu$-term. Thus, we must construct an algorithm for
calculating the function $B(\tan \beta,\text{sign}(\mu))$
and then find its roots at ${\Lambda_{\rm M}}$ by standard numerical methods.

The one-loop effective potential written in terms of the VEVs,
$v_u=\langle H_u^0 \rangle$ and $v_d=\langle H_d^0 \rangle$, is
\begin{equation}
V(Q)=V_0(Q)+\Delta V(Q) ,
\end{equation}
where
\begin{eqnarray}
V_0(Q) & = & \left( m_{H_d}^2+\mu^2 \right) v_d^2 + \left(
m_{H_u}^2+\mu^2 \right) v_u^2 -2 B v_u v_d+\frac 18 \left( g_L^2+g_Y^2
\right) \left(v_u^2-v_d^2 \right)^2 , \nonumber \\ \Delta V(Q) & = &
\frac 1{64 \pi^2} \sum_{k=\text{all the MSSM fields}} \left( -1
\right)^{2 S_k} n_k M_k^4 \left[ -\frac 32+\ln \frac{M_k^2}{Q^2}
\right] ,
\label{eq:potform}
\end{eqnarray}
where $S_k$, $n_k$, and $M_k$ are respectively the spin, the number of degrees 
of freedom, and the mass 
of the fields that contribute to $\Delta V$, and where all the couplings
are evaluated at some arbitrary renormalization scale $Q$. 

The full effective potential is independent of the renormalization
scale $Q$: the effects of changing the scale $Q$ in $\Delta V(Q)$ are
exactly cancelled as a result of the field and coupling constant 
renormalizations at the new scale.
However, the one-loop leading-logarithmic approximation we are using is most 
accurate when the 
logarithmic terms in (\ref{eq:potform}) are as small as possible, which is 
generally
true when $Q$ is chosen to equal $m_{\text{SUSY}} \simeq
m_{\text{squark}}$.

First we use the SM RGEs to evolve the
gauge and Yukawa couplings from the $m_{\rm Z}$ scale to the squark mass scale, 
$m_{\text{SUSY}}$. At 
the $m_{\text{SUSY}}$ scale we match the
SM and the MSSM parameters. This matching 
is straightforward at one-loop level, except for numerically
significant two-loop corrections $\delta_b$ to the down quark Yukawa
couplings in the high $\tan \beta$ regime~\cite{hallrattazzisarid}:
\begin{eqnarray}
g_a^{\text{MSSM}}(m_{\text{SUSY}}) & = &
g_a^{\text{SM}}(m_{\text{SUSY}}) , \nonumber \\
Y_{u}^{\text{MSSM}}(m_{\text{SUSY}}) & = &
Y_{u}^{\text{SM}}(m_{\text{SUSY}})/\sin \beta , \nonumber \\
Y_{d33}^{\text{MSSM}} (m_{\text{SUSY}}) & = &
Y_{d33}^{\text{SM}} (m_{\text{SUSY}})/\left( (1+\delta_b) \cos \beta
\right) , \nonumber \\ Y_{d,e}^{\text{MSSM}}(m_{\text{SUSY}}) & = &
Y_{d,e}^{\text{SM}}(m_{\text{SUSY}})/\cos \beta .
\label{eq:match}
\end{eqnarray}

At first iteration we simply ignore the two-loop threshold effects and
set $\delta_b=0$.  We use the MSSM RG equations to run the Yukawa
couplings to the messenger scale $\Lambda_{\rm M}$. 

At the messenger scale the values of the
mass-squared terms are given by the known two-loop graphs, while the
trilinear soft terms vanish. We then run the Yukawa couplings, the
mass-squared terms and the trilinear terms down to $m_{\text{SUSY}}$.

The minimization conditions for the tree-level potential $V_0$ are
\begin{equation}
\frac{\partial V_0}{\partial v_u} = \frac{\partial V_0}{\partial v_d}
=0 .
\label{eq:mincond1}
\end{equation}
Using (\ref{eq:mincond1}) one can solve the bilinear term, which
we denote by $B_0$, and the $\mu$-term, denoted by
$\mu_0=\text{sign}(\mu) \sqrt{\mu_0^2}$. As for $\delta_b$, it can be 
approximated  
by~\cite{hallrattazzisarid}
\begin{equation}
\delta_b = \mu \tan \beta \left[ \frac{2 \alpha_3}{3 \pi} M_3
I(m_{\tilde{b}_1}^2,m_{\tilde{b}_2}^2,M_3^2) + \frac{Y_{u33}}{16
\pi^2} A_{u33} Y_{u33}
I(m_{\tilde{t}_1}^2,m_{\tilde{t}_2}^2,\mu^2) \right] ,
\label{eq:deltab}
\end{equation}
where 
\begin{equation}
I(x,y,z)=-\frac{xy \ln x/y+yz \ln y/z +zx \ln z/x}{(x-y)(y-z)(z-x)} .
\end{equation}

At this stage we re-evaluate the down-Yukawa matrix at $m_{\text{SUSY}}$ using 
(\ref{eq:match}) and 
(\ref{eq:deltab}), and repeat the previous
steps to obtain the scalar mass-squared terms and the soft
trilinear couplings at scale $m_{\text{SUSY}}$.

We then calculate the mass eigenvalues of all the gauge bosons, fermions, 
and scalars. We check that all the scalar mass-squared eigenvalues are
non-negative, i.e., that we are at a true local minimum of the
tree-level potential. The obtained masses and their first derivatives with 
respect to the VEVs are then used to minimize the full one-loop effective
potential by using the equations
\begin{equation}
\frac{\partial V}{\partial v_u} = \frac{\partial V}{\partial v_d}
=0,
\label{eq:mineq0}
\end{equation}
from which we can solve $\mu=\text{sign}(\mu) \sqrt{\mu_0^2+\delta
\mu^2}$ and $B=B_0+\delta B$, where
\begin{eqnarray}
\delta\mu^2 & = & \frac 12 \frac{v_d \partial \Delta V /\partial
v_d-v_u \partial \Delta V/\partial v_u}{v_u^2-v_d^2} , \nonumber \\
\delta B & = & \frac 12 \frac{v_u \partial \Delta V /\partial v_d-v_d
\partial \Delta V/\partial v_u}{v_u^2-v_d^2} .
\label{eq:potcorr}
\end{eqnarray}

Numerically, the most significant one-loop contribution to $\delta B$
and $\delta \mu^2$ comes from the (s)top and (s)bottom loops. The
contribution can be obtained by a straightforward calculation from
eqs. (\ref{eq:potform}) and (\ref{eq:potcorr}) together with the quark
masses $m_t=\lambda_t v_u$ and $m_b=\lambda_b v_d$ and the squark mass
matrices.

Now all the parameters at the scale $m_{\text{SUSY}}$ are known. We then
finally run $B$ and $\mu$ from $m_{\text{SUSY}}$ to $\Lambda_{\rm M}$.
It remains to use numerical iteration to find the values of $\tan
\beta$ and $\text{sign}(\mu)$ that satisfy $B_{\Lambda_{\rm M}}(\tan
\beta,\text{sign}(\mu))=0$, where we denote $B$ at the scale
 $\Lambda_{\rm M}$ by $B_{\Lambda_{\rm M}}$.

We use one-loop RGEs for all MSSM parameters, except for the bilinear
scalar coupling $B$, for which we use the full two-loop
expression~\cite{martinvaughn}. Including the two-loop contribution to
the running of $B$-parameter decreases (increases) $\tan \beta$
prediction for positive (negative) $\mu$-parameter 
(for a discussion on including the two-loop RGE
running see~\cite{borzu}).


We have listed all models consistent with radiative symmetry breaking
in table~\ref{tab:modellist}, taking $\Lambda_{\text{SUSY}} =100$ TeV
and $\Lambda_{\rm M}=100\, \Lambda_{\text{SUSY}}$.
There is a total of 32 possible messenger multiplicities and for each
messenger multiplicity there can be several possible solutions
corresponding to different roots of the function
$B_{\Lambda_{\rm M}}(\tan \beta,\text{sign}(\mu))$. 

\begin{figure}
\caption{The values of the bilinear term $B_{\Lambda_{\rm M}}$ at the
messenger scale as a function of $\tan \beta$. The thick lines
correspond to the minimal model with messenger multiplicities
$(0,0,1,1,0)$ at $\Lambda_{\text{SUSY}}=100$ TeV and the thin lines
correspond to the model with messenger multiplicities $(0,0,1,2,0)$ at
$\Lambda_{\text{SUSY}}=150$ TeV. The solid lines are for
${\text{sign}}(\mu)=+1$ and dashed lines are for
${\text{sign}}(\mu)=-1$. The minimal model has only one root
consistent with the vanishing bilinear term at the messenger scale,
while the other model has three, as can also be seen in figure 2. The
lines are plotted up to the values of $\tan \beta$ below which the
radiative symmetry breaking leads to the $SU(3)_C \times U(1)_{em}$
preserving physical vacuum state.}
\label{fig:roots}
\mbox{\epsfxsize=16cm \epsfysize=16cm \epsffile{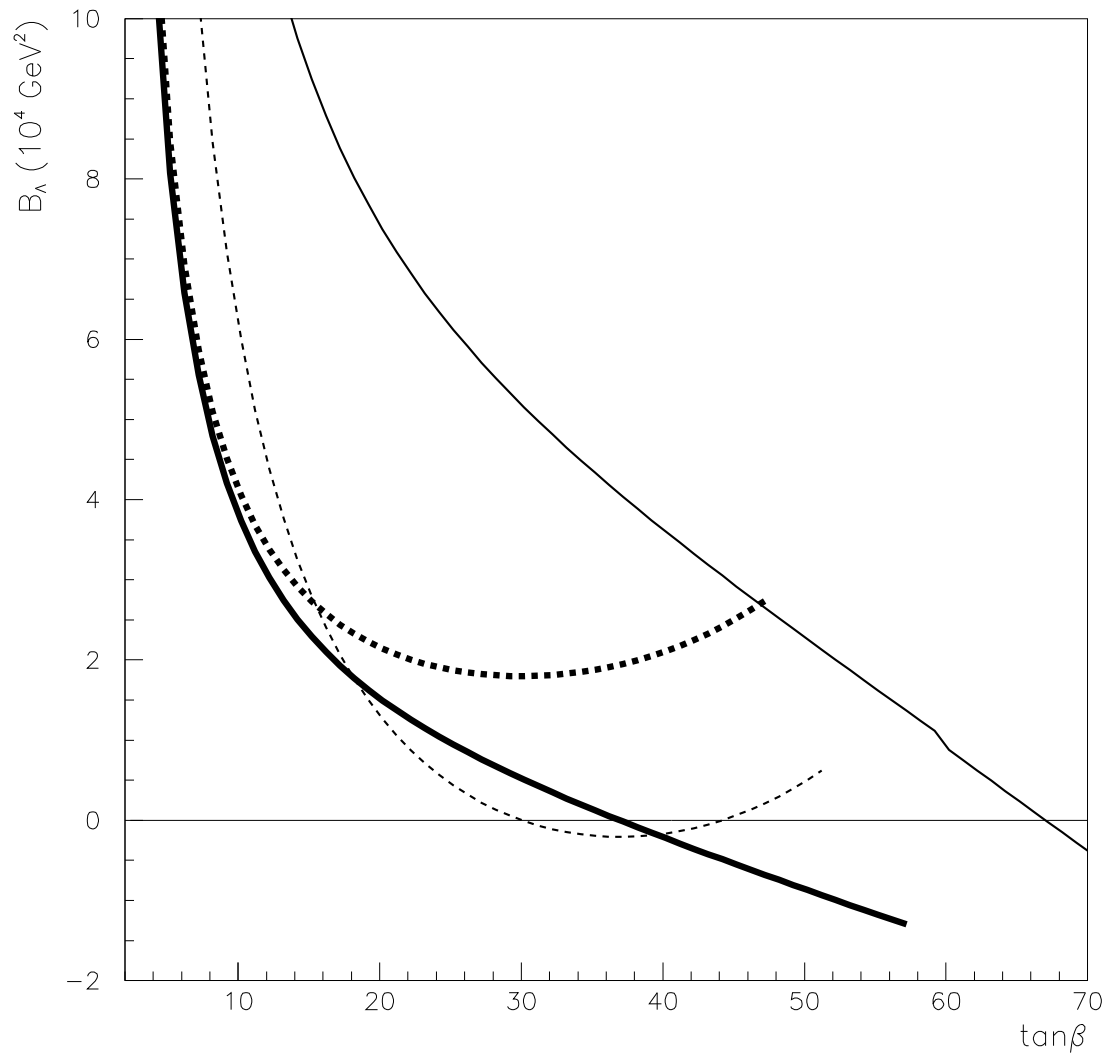}}
\end{figure}

In Fig.~\ref{fig:roots} we have plotted the function
$B_{\Lambda_{\rm M}}(\tan \beta,\text{sign}(\mu))$ for some representative
models.  In the minimal case $(0,0,1,1,0)$ there is only one root
corresponding to a positive $\mu$-term. There are three possible models
having messenger multiplicities $(0,0,1,2,0)$, one root corresponding
to positive $\mu$-terms and two roots corresponding to a negative $\mu$-term.
For all messenger configurations the corresponding lines are cut above
certain values of $\tan \beta$, since they would not lead to the
desired radiative symmetry breaking: either the $SU(3)_C \times
U(1)_{em}$ preserving vacuum state would actually be a saddle point of
the effective potential due to the negative stau mass-squared
eigenvalue, or the minimization conditions in eq. (\ref{eq:mineq0})
would have no solutions for $|\mu^2| \ge 0$.

For every model we have calculated the ratio of the branching ratio
$BR(b \rightarrow s \gamma)$ by using the full one-loop expression
given in~\cite{bertolini}. We find that the most important non-SM
contribution comes from the constructive interference with the charged
Higgs loop. For a positive (negative) $\mu$-term the contribution from
the chargino channel is typically approximately 5-30 \% destructive
(constructive). The gluino amplitude is typically about one per cent
of the SM amplitude, while the neutralino contribution turns out to be
totally negligible.

We find that in all the representative cases, except some models with
very low SUSY breaking scale, the branching ratio $BR(b\rightarrow s
\gamma)$ is larger than the SM prediction. The experimental bounds
($1.0<10^4\times BR_{\rm EXP}(b \rightarrow s \gamma ) < 4.2$
\cite{cleo}) combined with the theoretical uncertainty in the SM
prediction ($10^4\times BR_{\rm SM}(b \rightarrow s \gamma )=3.5 \pm
0.3$ \cite{buras}) limit the branching ratio to be between 0.3 and 1.4
times the SM prediction.  At $\Lambda_{\text{SUSY}}=100$ TeV, all
models obey this limit.

\begin{figure}
\caption{The ratio $\Gamma (b \rightarrow s \gamma) / \Gamma_{SM}$ as
a function of the scale $\Lambda_{\text{SUSY}}$ ($\Lambda_{\rm
M}=100\, \Lambda_{\text{SUSY}}$). The solid lines correspond to the
models with ${\text{sign}} (\mu )=+1$ and the dashed lines correspond
to the models with ${\text{sign}} (\mu )=-1$. At
$\Lambda_{\text{SUSY}}=100$ TeV these lines correspond to the models
listed in table~I.  Models with large rates are identified by
$(n_Q,n_U,n_D,n_L,n_E)$ multiplicities.}
\label{fig:bsg}
\mbox{\epsfxsize=16cm \epsfysize=16cm \epsffile{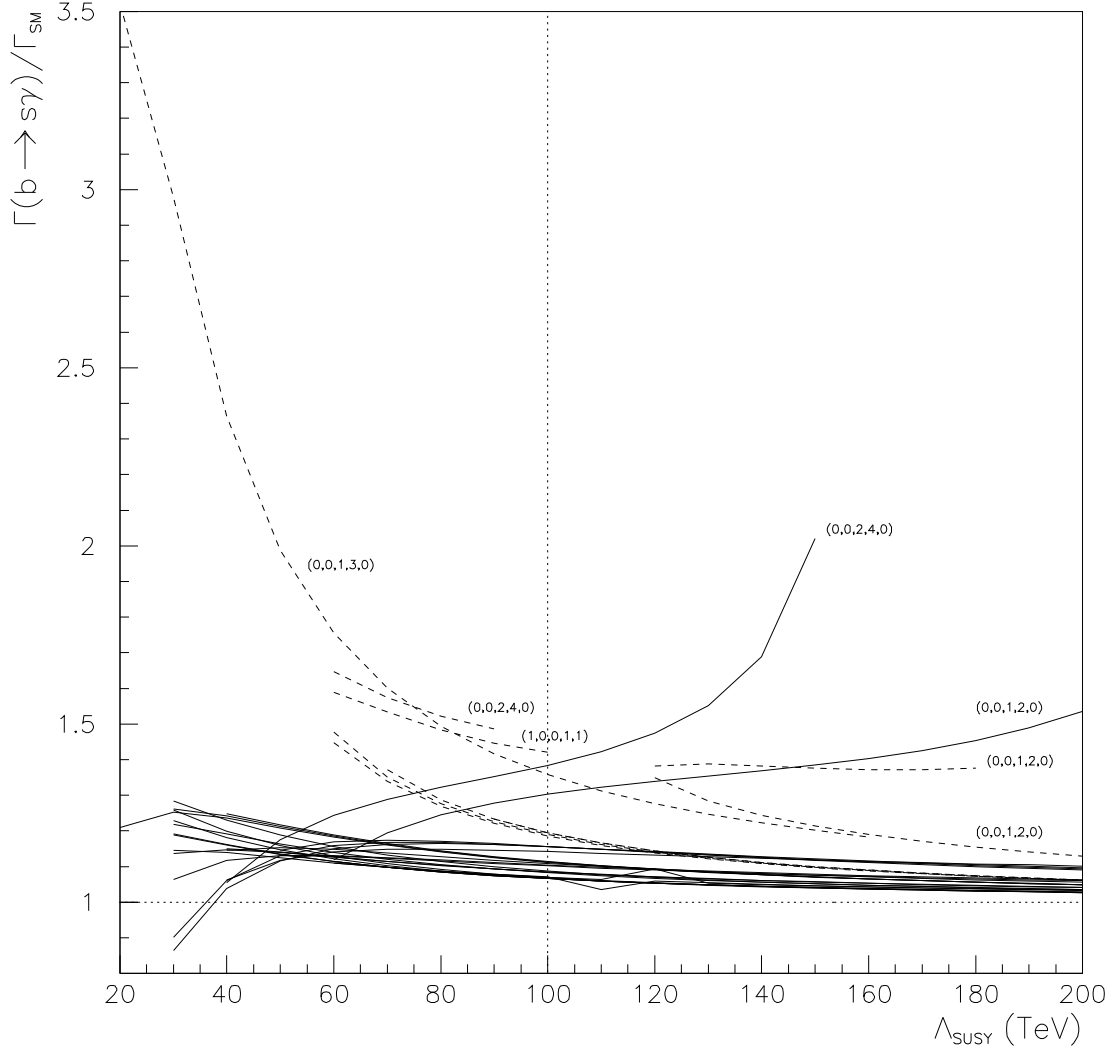}}
\end{figure}

In Fig.~\ref{fig:bsg} we have plotted the branching ratios
$BR(b\rightarrow s \gamma)$ for different messenger multiplicities and
for different values of $\Lambda_{\text{SUSY}}$. The sparticle masses,
which are shown in table~\ref{tab:modellist} for
$\Lambda_{\text{SUSY}}=100$ TeV, are roughly proportional to the scale
$\Lambda_{\text{SUSY}}$. For ${\text{sign}}(\mu)=+1$ the SM channel
and the Higgs channel interfere destructively with the chargino
channel. This interference keeps the branching ratio $BR(b\rightarrow
s \gamma)$ approximately constant for all scales for models with heavy
charged Higgs fields. In the case ${\text{sign}}(\mu)=-1$ the chargino
amplitude interferes constructively. As a result, the branching ratio
can grow unacceptably large for small SUSY breaking scales.

To summarize, we have described in detail an algorithm to calculate
all the unmeasured parameters of the MSSM for a given messenger
sector with vanishing bilinear and trilinear soft terms. 
We have performed this calculation for a class of generalized
GMSB models, giving the full particle spectrum for each model. 

We have also calculated the branching ratios $BR(b\rightarrow s \gamma
)$ for these models. We find that the branching ratio is almost always
greater than the SM prediction. For most of these models the predicted
branching ratios are less than 1.3 times the SM prediction. However,
models with light ($m_H \alt 300$ GeV) charged Higgs boson, or small
supersymmetry breaking scales and negative $\mu$-parameter, result in 
unacceptably large $BR(b\rightarrow s \gamma )$.

\section*{Acknowledgments}

HH thanks the Swedish Natural Science Research Council for financial
support.  The work of KH, KP and DXZ is partially supported by the Academy of
Finland (No.~37599).

We would like to thank F. Borzumati and T. Kobayashi for useful discussions 
and correspondence.



\begin{table}
\caption{
The MSSM particle spectrum for various messenger configurations 
($\Lambda_{\text{SUSY}}=100\, \text{TeV}=\Lambda_{\rm
M}/100$). The values of $\tan \beta$ and $\mu$ are calculated using
one-loop effective potential, as described in text. The masses of the
charged Higgs boson and spartners, expressed in GeV, are derived from
the tree-level potential at the squark mass scale. At high $\tan
\beta$ regime the lighter scalar Higgs boson mass is, after radiative
corrections, about 130 GeV, while the pseudoscalar, charged and heavier
scalar Higgs bosons are approximately mass-degenerate.}
\label{tab:modellist}
\begin{tabular}{lrrrrrrr}
$(n_Q,n_U,n_D,n_L,n_E)$ & $\tan \beta$ & $\mu$ & $m_{H^\pm}$ &
$m_{{\tilde{\chi}}^\pm_{1,2}}$ & $m_{{\tilde{\chi}}^0_{1,2,3,4}}$ &
$m_{{\tilde{e}}_{1,2}}$ & $m_{{\tilde{\tau}}_{1,2}}$ \\ & $\frac{\Gamma (b
\rightarrow s \gamma)}{\Gamma_{SM}}$ & $M_3$ &
$m_{{\tilde{\nu}}_e}/m_{{\tilde{\nu}}_{\tau}}$ & $m_{{\tilde{u}}_{1,2}}$ &
$m_{{\tilde{t}}_{1,2}}$ & $m_{{\tilde{d}}_{1,2}}$ &
$m_{{\tilde{b}}_{1,2}}$ \\ 
\tableline $(0,0,1,1,0)$ & $36$ & $504$ &
 $511$ &
 $248/503$ &
 $136/248/491/502$ & $192/362$ &
 $141/370$ \\
 & $1.2$ & $703$ & $353/348$ & $953/1006$ &
 $821/945$ & $947/1009$ & $895/943$ \\
\tableline $(0,0,1,1,1)$ & $39$ & $499$ &
 $503$ &
 $248/497$ &
 $247/301/485/499$ & $288/377$ &
 $241/389$ \\
 & $1.2$ & $703$ & $369/362$ & $963/1006$ &
 $830/943$ & $950/1010$ & $892/941$ \\
\tableline $(0,0,1,1,2)$ & $41$ & $493$ &
 $495$ &
 $248/492$ &
 $247/447/479/512$ & $365/394$ &
 $309/414$ \\
 & $1.2$ & $703$ & $386/377$ & $975/1007$ &
 $839/941$ & $953/1010$ & $890/940$ \\
\tableline $(0,0,1,2,0)$ & $63$ & $399$ &
 $285$ &
 $356/538$ &
 $216/359/377/538$ & $243/526$ &
 $105/513$ \\
 & $1.3$ & $698$ & $520/500$ & $954/1070$ &
 $804/967$ & $945/1073$ & $820/958$ \\
\tableline $(0,0,1,3,0)$ & $27$ & $-223$ &
 $599$ &
 $168/781$ &
 $159/175/313/781$ & $287/664$ &
 $260/659$ \\
 & $1.4$ & $695$ & $660/654$ & $956/1139$ &
 $786/1066$ & $943/1142$ & $905/1055$ \\
\tableline $(0,0,2,1,0)$ & $21$ & $788$ &
 $804$ &
 $253/775$ &
 $193/253/769/774$ & $226/364$ &
 $198/375$ \\
 & $1.1$ & $1364$ & $356/354$ & $1455/1488$ &
 $1279/1414$ & $1450/1490$ & $1394/1438$ \\
\tableline $(0,0,2,1,1)$ & $22$ & $785$ &
 $802$ &
 $253/771$ &
 $253/359/765/771$ & $313/380$ &
 $282/397$ \\
 & $1.1$ & $1364$ & $372/369$ & $1462/1488$ &
 $1285/1413$ & $1452/1490$ & $1392/1438$ \\
\tableline $(0,0,2,1,2)$ & $23$ & $781$ &
 $801$ &
 $253/767$ &
 $253/524/761/769$ & $387/397$ &
 $343/427$ \\
 & $1.1$ & $1364$ & $389/386$ & $1470/1489$ &
 $1292/1413$ & $1454/1491$ & $1391/1439$ \\
\tableline $(0,0,2,2,0)$ & $38$ & $724$ &
 $726$ &
 $499/716$ &
 $276/499/701/716$ & $272/525$ &
 $218/528$ \\
 & $1.1$ & $1359$ & $519/511$ & $1454/1530$ &
 $1269/1433$ & $1447/1532$ & $1381/1431$ \\
\tableline $(0,0,2,3,0)$ & $55$ & $644$ &
 $569$ &
 $596/790$ &
 $358/598/615/790$ & $313/662$ &
 $202/651$ \\
 & $1.1$ & $1355$ & $657/639$ & $1454/1577$ &
 $1256/1449$ & $1444/1579$ & $1326/1443$ \\
\tableline $(0,0,2,4,0)$ & $66$ & $537$ &
 $297$ &
 $494/1033$ &
 $430/501/508/1033$ & $351/787$ &
 $160/761$ \\
 & $1.4$ & $1352$ & $783/753$ & $1455/1628$ &
 $1243/1469$ & $1441/1630$ & $1257/1460$ \\
\tableline $(0,0,2,4,0)$ & $23$ & $-539$ &
 $852$ &
 $499/1033$ &
 $432/505/511/1033$ & $351/787$ &
 $330/783$ \\
 & $1.2$ & $1352$ & $783/778$ & $1455/1628$ &
 $1244/1533$ & $1441/1630$ & $1408/1524$ \\
\tableline $(0,0,3,1,0)$ & $16$ & $1015$ &
 $1025$ &
 $253/998$ &
 $249/254/994/997$ & $257/368$ &
 $232/381$ \\
 & $1.1$ & $2009$ & $359/358$ & $1902/1925$ &
 $1691/1836$ & $1897/1927$ & $1819/1886$ \\
\tableline $(0,0,3,2,0)$ & $26$ & $965$ &
 $987$ &
 $506/948$ &
 $333/506/941/948$ & $299/526$ &
 $267/532$ \\
 & $1.1$ & $2005$ & $520/516$ & $1900/1956$ &
 $1683/1853$ & $1893/1958$ & $1832/1870$ \\
\tableline $(0,0,3,3,0)$ & $39$ & $905$ &
 $898$ &
 $745/902$ &
 $416/745/878/902$ & $338/661$ &
 $278/660$ \\
 & $1.1$ & $2001$ & $656/647$ & $1899/1992$ &
 $1674/1867$ & $1890/1993$ & $1811/1865$ \\
\tableline $(0,0,3,4,0)$ & $52$ & $833$ &
 $759$ &
 $788/1038$ &
 $499/790/802/1038$ & $374/785$ &
 $268/773$ \\
 & $1.1$ & $1997$ & $781/762$ & $1898/2031$ &
 $1663/1881$ & $1887/2033$ & $1762/1875$ \\
\tableline $(0,0,4,1,0)$ & $14$ & $1216$ &
 $1222$ &
 $253/1199$ &
 $253/306/1196/1198$ & $285/372$ &
 $259/388$ \\
 & $1.1$ & $2645$ & $363/362$ & $2322/2339$ &
 $2081/2236$ & $2316/2341$ & $2220/2307$ \\
\tableline $(0,0,4,2,0)$ & $21$ & $1173$ &
 $1199$ &
 $507/1155$ &
 $389/507/1150/1155$ & $325/528$ &
 $300/535$ \\
 & $1.1$ & $2641$ & $522/520$ & $2320/2364$ &
 $2075/2251$ & $2313/2365$ & $2235/2293$ \\
\tableline $(0,0,4,3,0)$ & $30$ & $1123$ &
 $1143$ &
 $758/1108$ &
 $473/758/1099/1108$ & $362/662$ &
 $322/664$ \\
 & $1.1$ & $2637$ & $657/651$ & $2318/2392$ &
 $2067/2264$ & $2309/2394$ & $2239/2275$ \\
\tableline $(0,0,4,4,0)$ & $41$ & $1066$ &
 $1047$ &
 $973/1087$ &
 $556/974/1039/1088$ & $396/784$ &
 $329/780$ \\
 & $1.1$ & $2633$ & $780/768$ & $2317/2424$ &
 $2058/2275$ & $2306/2425$ & $2213/2273$ \\
\tableline $(1,0,0,0,1)$ & $54$ & $648$ &
 $582$ &
 $600/790$ &
 $193/600/620/790$ & $226/653$ &
 $29/644$ \\
 & $1.1$ & $1355$ & $648/632$ & $1447/1576$ &
 $1250/1452$ & $1442/1578$ & $1329/1445$ \\
\tableline $(1,0,0,1,0)$ & $24$ & $-548$ &
 $832$ &
 $509/1033$ &
 $110/509/516/1033$ & $171/772$ &
 $122/768$ \\
 & $1.2$ & $1352$ & $768/763$ & $1440/1627$ &
 $1231/1531$ & $1438/1629$ & $1399/1522$ \\
\tableline $(1,0,0,1,1)$ & $23$ & $-544$ &
 $843$ &
 $504/1033$ &
 $275/506/511/1033$ & $272/779$ &
 $244/775$ \\
 & $1.2$ & $1352$ & $775/770$ & $1447/1628$ &
 $1237/1532$ & $1439/1630$ & $1404/1523$ \\
\tableline $(1,0,0,1,1)$ & $45$ & $-542$ &
 $478$ &
 $501/1033$ &
 $275/503/508/1033$ & $272/779$ &
 $173/767$ \\
 & $1.4$ & $1352$ & $775/762$ & $1447/1628$ &
 $1237/1482$ & $1439/1630$ & $1287/1474$ \\
\tableline $(1,0,1,0,1)$ & $38$ & $908$ &
 $905$ &
 $745/905$ &
 $250/745/882/904$ & $257/652$ &
 $184/651$ \\
 & $1.1$ & $2001$ & $647/638$ & $1894/1991$ &
 $1669/1869$ & $1888/1993$ & $1813/1867$ \\
\tableline $(1,0,1,1,1)$ & $51$ & $837$ &
 $770$ &
 $792/1038$ &
 $333/792/806/1038$ & $299/777$ &
 $167/766$ \\
 & $1.1$ & $1997$ & $773/756$ & $1893/2031$ &
 $1658/1882$ & $1885/2032$ & $1765/1877$ \\
\tableline $(1,0,2,0,0)$ & $29$ & $1128$ &
 $1152$ &
 $758/1113$ &
 $139/758/1104/1113$ & $190/643$ &
 $126/645$ \\
 & $1.1$ & $2637$ & $639/634$ & $2309/2392$ &
 $2059/2266$ & $2307/2393$ & $2242/2276$ \\
\tableline $(1,0,2,0,1)$ & $29$ & $1126$ &
 $1147$ &
 $758/1111$ &
 $306/758/1102/1110$ & $285/652$ &
 $241/654$ \\
 & $1.1$ & $2637$ & $647/642$ & $2314/2392$ &
 $2063/2265$ & $2308/2393$ & $2241/2276$ \\
\tableline $(1,0,2,1,0)$ & $39$ & $1071$ &
 $1062$ &
 $976/1091$ &
 $222/976/1045/1091$ & $241/768$ &
 $137/765$ \\
 & $1.1$ & $2633$ & $764/754$ & $2307/2424$ &
 $2050/2278$ & $2304/2425$ & $2217/2276$ \\
\tableline $(1,0,2,1,1)$ & $40$ & $1069$ &
 $1054$ &
 $974/1089$ &
 $390/975/1042/1089$ & $325/776$ &
 $248/772$ \\
 & $1.1$ & $2633$ & $772/761$ & $2312/2424$ &
 $2053/2277$ & $2305/2425$ & $2215/2274$ 
\end{tabular}
\end{table}

\end{document}